\documentclass[revtex,aps,prb,reprint,superscriptaddress,amsmath,amssymb,showkeys,showpacs,preprintnumbers]{revtex4-1}

\usepackage{graphicx}
\usepackage{dcolumn}
\usepackage{bm}
\usepackage{float}

\begin{document}

\title{Spin-current absorption by inhomogeneous spin-orbit coupling}

\author{Kazuhiro Tsutsui}
\email[]{tsutsui@stat.phys.titech.ac.jp}
\affiliation{Department of Physics, Tokyo Institute of Technology, 2-12-1 Ookayama, Meguro-ku, Tokyo 152-8551, Japan}

\author{Kazuhiro Hosono}
\affiliation{
International Center of Materials Nanoarchitectonics (WPI-MANA), Namiki 1-1, Tsukuba 305-0044, Japan}

\author{Takehito Yokoyama}
\affiliation{Department of Physics, Tokyo Institute of Technology, 2-12-1 Ookayama, Meguro-ku, Tokyo 152-8551, Japan}

\date{\today}

\begin{abstract}
We investigate the spin-current absorption induced by an inhomogeneous spin-orbit coupling due to impurities in metals.
We  consider the system with spin currents driven by the electric field or the spin accumulation.
The resulting diffusive spin currents, including the gradient of the spin-orbit coupling strength, indicate the spin-current absorption at the interface, which is exemplified with experimentally relevant setups. 
\end{abstract}


\maketitle

\section{Introduction}

Spintronics aims to utilize not only the charge degree of freedom but also the spin degree of freedom \cite{Zutic}.
Spin currents are an important notion in spintronics from the aspect of both basic and applied science.
Spin currents are generated by a current injection into a ferromagnet or spin pumping using the magnetization precession\cite{Silsbee,Tserkovnyak}.
Other methods to generate spin currents are the spin Hall effect \cite{Dyakonov71, Hirsch99, Murakami03} or the excitation of the spin wave \cite{Kajiwara10}.
In the development of spintronic devices, techniques to detect spin currents efficiently are indispensable as well as the spin-current generation.
For instance, spin currents have been successfully detected via the inverse spin Hall effect \cite{Saitoh06, Valenzuela06, Kimura07}.
Due to this effect, spin currents are converted into electric currents and therefore are observed as a voltage drop.
The electric voltage induced by the spin current has been observed in a lateral junction of a ferromagnetic metal and a nonmagnetic metal with spin-orbit coupling (SOC) such as aluminum \cite{Valenzuela06} or platinum \cite{Saitoh06, Kimura07}.
In particular, platinum is a typical spin-current detector because of its strong SOC due to impurities, which is used to absorb a spin current \cite{Kajiwara10}.

Kimura \textit{et al.} demonstrated the spin-current detection via the inverse spin Hall effect and suggested that the absorption of the spin current occurs from the Cu cross into the Pt wire \cite{Kimura07}.
They explained the spin-current absorption using the resistance mismatch for the spin current.
Namely, they defined the spin resistance as $R_s = \lambda/[\sigma S (1-P^2)]$ with the spin diffusion length $\lambda$, the spin polarization $P$, the conductivity $\sigma$ and the effective cross-sectional area $S$ for the spin current, and then considered that the spin current flowing in a material is absorbed into the adjacent material with smaller spin resistance \cite{Kimura07, Kimura08}.
Since platinum is a strong spin-orbit coupled material while copper is a weak one,  the spin resistance of Pt is smaller than that of Cu and hence the injected spin current in the Cu cross is partly absorbed into the Pt wire.
Here, the difference of the SOC strength, i.e., the inhomogeneous SOC is the key to the absorption of spin current.
As shown above, the physics of the spin-current absorption induced by an inhomogeneous SOC has been understood only phenomenologically,
in that previous theories involve the phenomenological parameters such as the spin-diffusion length or the spin resistance
\cite{Kimura05, Takahashi03} and microscopic theories of the spin-current absorption are missing.
The spin Hall effect \cite{Tserkovnyak07, Dugaev10, Glazov10} and the spin-current generation \cite{Tsutsui11} due to an inhomogeneous Rashba SOC have been theoretically examined.
Also, a spin current generation near an interface due to interfacial spin-orbit coupling has been investigated\cite{Linder}.

In the present work, we consider the spin-current absorption as follows.
In general, the spin-continuity equation for spin-orbit coupled systems has a source term: $\frac{\partial s^\alpha}{\partial t}+\nabla\cdot \bm{j}_s^\alpha={\cal T}^\alpha$.
Here, $s^\alpha$, $\bm{j}_s^\alpha$ and ${\cal T}^\alpha$ represent the spin density, the spin current and the spin torque (or source term) with $\alpha$ being a component in spin space, respectively.
Thus, the divergence of spin currents in the steady state is non-zero and therefore leads to the generation of spin currents.
We refer to a spin-current generation induced near an interface between different spin-orbit coupled materials or due to an inhomogeneous SOC as the spin-current absorption.
On the other hand, the spin-current absorption in previous phenomenological studies is based on the continuity of a spin current at an interface, and therefore absorbed spin currents are not generated ones at the interface.

In this paper, we theoretically examine the spin-current generation induced by an inhomogeneous SOC due to impurities.
We consider the spin-current absorption in systems where spin currents are generated by an external electric field in Sec. II.
In Sec. III, we also investigate systems where spin currents are induced by spin accumulation.
We present analytical expressions of the spin currents as a response to the gradient of the SOC strength.
Then, we apply these results to the vicinity of the interface between metals with different SOC strengths, verifying the absorption of spin currents. 
Section IV is devoted to the discussions. We summarize the paper in Sec. V. 


\section{Absorption of spin current driven by external field}

\subsection{Model and Formalism}

We consider the ferromagnetic conductor in the presence of the spin-orbit scattering due to impurities \cite{Tse06, Dugaev01, Dugaev05, Hosono} and assume that the coupling constant of the spin-orbit scattering slowly varies in space.
This spatial variation of the SOC has not been considered so far.
In order to address the spin-current absorption, we describe the input spin current by the spin-polarized current flowing in a ferromagnet under an external electric field.
The total Hamiltonian is thus composed of the free-electron part with the exchange coupling (${\cal H}_{\mathrm{FM}}$), the impurity-scattering part (${\cal H}_{\mathrm{imp}}$), the SOC due to impurities with its minimal substitution (${\cal H}^0_{\mathrm{SO}}+{\cal H}^A_{\mathrm{SO}}$), and the interaction between the vector potential and the electric current (${\cal H}_{\mathrm{em}}$):
\begin{eqnarray}
{\cal H}_{\mathrm{FM}} &=& \sum_{\sigma=\pm 1} \int d\bm{r}\ c_\sigma^\dagger(\bm{r},t)\left(-\frac{\hbar^2}{2m}\nabla^2-\epsilon_{F\sigma} \right)c_\sigma(\bm{r},t), \nonumber \\
\\
{\cal H}_{\mathrm{imp}} &=&  \sum_{\sigma=\pm 1} \int d\bm{r}\ U(\bm{r})\ c_\sigma^\dagger(\bm{r},t)c_\sigma(\bm{r},t), \\
{\cal H}^0_{\mathrm{SO}} &=& \frac{\hbar}{2i} \sum_{\sigma,\sigma'=\pm1} \int d\bm{r}\ \lambda_{\mathrm{SO}}(\bm{r}) \nonumber \\
 & & \times \nabla U(\bm{r}) \cdot c_\sigma^\dagger(\bm{r},t) \left( \overleftrightarrow{\nabla}\times \bm{\sigma}_{\sigma \sigma'} \right) c_{\sigma'} (\bm{r},t), \\
{\cal H}^A_{\mathrm{SO}} &=& e\sum_{\sigma,\sigma'=\pm1} \int d\bm{r}\ \lambda_{\mathrm{SO}}(\bm{r}) \nonumber \\
& & \times \nabla U(\bm{r})\cdot c_\sigma^\dagger(\bm{r},t) (\bm{A}(\bm{r},t)\times \bm{\sigma}_{\sigma \sigma'}) c_{\sigma'} (\bm{r},t), \\
{\cal H}_{\mathrm{em}} &=& \frac{e\hbar}{2mi} \sum_{\sigma=\pm 1} \int d\bm{r}\ \bm{A}(\bm{r},t)\cdot c_\sigma^\dagger(\bm{r},t)\overleftrightarrow{\nabla} c_\sigma(\bm{r},t).
\end{eqnarray}
Here, $c_\sigma(\bm{r},t)$ ($c^\dagger_\sigma(\bm{r},t)$) represents the annihilation (creation) operator of a conduction electron, $\bm{\sigma}_{\alpha\beta}$ represents Pauli matrices, $e (>0)$ is the electric charge and $c^\dagger_\sigma \overleftrightarrow{\nabla} c_\sigma \equiv c^\dagger_\sigma \nabla c_\sigma -(\nabla c^\dagger_\sigma) c_\sigma$.
$\sigma=1$ and $\sigma=-1$ correspond to the up ($\uparrow$) and down spins ($\downarrow$), respectively.
We set $\epsilon_{F\sigma}\equiv \epsilon_F+\sigma M$ with $\epsilon_F$ the Fermi energy and $M$ the exchange energy.
$\lambda_{\mathrm{SO}}(\bm{r})$ represents the SOC strength with spatial variation, averaged over the impurity positions. $\bm{A}(\bm{r},t)$ is the vector potential for the external electric field and we adopt the fixed gauge as $\bm{E}_{\mathrm{em}}=-\dot{\bm{A}}$.
$U(\bm{r})$ is the short ranged random impurity potential and averaging over the impurity positions is carried out as $\langle U(\bm{r}) U(\bm{r}')\rangle_{\mathrm{imp}}=u_0^2 n_{\mathrm{imp}} \delta(\bm{r}-\bm{r}')$, where $u_0$ and $n_{\mathrm{imp}}$ are the strength of the impurity potential and the impurity concentration, respectively.
Here, we assume that the exchange field is much stronger than the stray field and hence we neglect the effect of the stray (magnetic) field in our model.
Note that, in the Hamiltonian ${\cal H}_{\mathrm{SO}}^0$, the operator $\nabla$ does not act on $\lambda_{\mathrm{SO}}(\bm{r})$. 

\begin{figure}
 \begin{center} 
 \includegraphics[width=80mm]{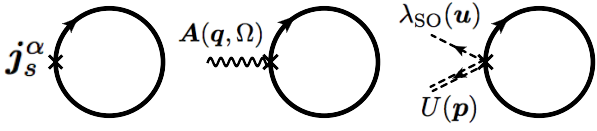} 
 \end{center} 
 \caption{Diagrams of the spin current for the total Hamiltonian. Bold lines denote the total Green's function. An wavy line denotes the vector potential ($\bm{A}(\bm{q},\Omega)$). A broken line and double broken line represent the SOC strength ($\lambda_{\mathrm{SO}}(\bm{u})$) and the impurity potential ($U(\bm{p})$), respectively.} 
 \label{spinc} 
 \end{figure}

The quantum description of the spin current is obtained by identifying the Heisenberg equation of motion for the spin density with the spin-continuity equation \cite{Tatara08, Hosono}.
The spin current operator in the $i$-direction with $\alpha$-spin polarization reads
\begin{eqnarray}
(\hat{j}_s^\alpha)_i\equiv \frac{1}{2m}\frac{\hbar}{i} c^\dagger(\bm{r},t) \sigma^\alpha \overleftrightarrow{\partial}_i c(\bm{r},t) -\frac{e}{m} A_i \sigma^\alpha c^\dagger(\bm{r},t) c(\bm{r},t) \nonumber \\
+\lambda_{\mathrm{SO}} \sum_j \epsilon_{\alpha j i} \partial_j U (\bm{r}) c^\dagger(\bm{r},t) c(\bm{r},t). \nonumber \\
& &
\end{eqnarray}
The spin current is thus given by (see also Fig. \ref{spinc} )
\begin{eqnarray}
(j_s^\alpha)_i &=& \frac{1}{2m}\frac{\hbar^2}{i} \sum_{\bm{k},\bm{k}'} e^{i(\bm{k}-\bm{k}')\cdot \bm{r}} (k+k')_i \mathrm{tr} \left[ \sigma^\alpha G_{\bm{k},\bm{k}'}(t) \right]^< \nonumber
\end{eqnarray}
\begin{eqnarray}
 & & -\frac{e\hbar}{m} \int \frac{d\Omega}{2\pi} e^{i\Omega t} \sum_{\bm{k},\bm{k}',\bm{q}} e^{i(\bm{k}-\bm{k}'+\bm{q})\cdot \bm{r}} A_i(\bm{q},\Omega) \mathrm{tr} \left[ \sigma^\alpha G_{\bm{k},\bm{k}'}(t) \right]^< \nonumber \\
 & & +i\hbar \sum_j \sum_{\bm{k},\bm{k}',\bm{u},\bm{p}} e^{i(\bm{k}-\bm{k}'+\bm{u}+\bm{p})\cdot \bm{r}} \epsilon_{\alpha j i} p_j \lambda_{\mathrm{SO}}(\bm{u}) U(\bm{p}) \mathrm{tr} \left[ G_{\bm{k},\bm{k}'}(t) \right]^<. \nonumber \\
 & & \label{sc3tp}
\end{eqnarray}
Here, $G_{\bm{k},\bm{k}'}(t,t') \equiv \frac{1}{i\hbar}\langle T_c\ [c_{\bm{k}}(t) c^\dagger_{\bm{k}'}(t')] \rangle$ denotes the time-ordered Green's function (Keldysh Green's function) of the total Hamiltonian and $G_{\bm{k},\bm{k}'}(t) \equiv \lim_{t'\to t} G_{\bm{k},\bm{k}'}(t,t')$.
$[\cdots ]^<$ means taking the lesser component of the Green's functions and $\mathrm{tr} [\cdots]$ means taking trace over spin.

We perturbatively calculate the spin currents induced by the inhomogeneity of the SOC strength using the Keldysh Green's function formalism.
We consider ${\cal H}_{\mathrm{FM}}+{\cal H}_{\mathrm{imp}}$ as the non-perturbative part and ${\cal H}^0_{\mathrm{SO}},$ ${\cal H}^A_{\mathrm{SO}}$ and ${\cal H}_{\mathrm{em}}$ as perturbative parts.
We remark that the non-perturbative Green's function contains the self energy by the impurity scattering within the first Born approximation.
Within the linear response with respect to the electric field, the second part of the spin currents in Eq.(\ref{sc3tp}) corresponds to an equilibrium spin current, and therefore we ignore this contribution.
The leading contributions of the spin current contain the first order term with respect to the SOC strength $\lambda_{\mathrm{SO}}$.
Since we are interested in spin currents generated by the inhomogeneous SOC strength, we focus on the contribution with the spatial derivative of the SOC strength.
The third part of the spin currents, on the other hand, does not involve the spatial derivative of the SOC strength.
Consequently, the first part of Eq.(\ref{sc3tp}) is relevant to our study.

\subsection{Local spin current}

\begin{figure}
 \begin{center} 
 \includegraphics[width=85mm]{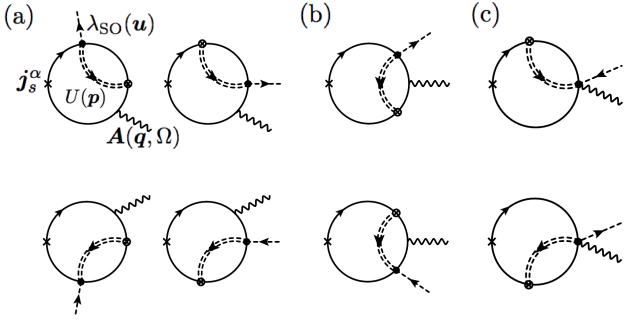} 
 \end{center} 
 \caption{Diagrams of the leading contribution of the local spin current induced by the inhomogeneity of the SOC strength.
 Solid lines denote the free Green's function.
 Broken lines denote the SOC strength ($\lambda_{\mathrm{SO}}(\bm{u})$) and double broken lines denote the impurity potential ($U(\bm{p})$).
 Wavy lines represent the vector potential ($A_l(\bm{q},\Omega)$).} 
 \label{FD} 
 \end{figure}
We first treat the local spin current, which is driven by the external electric field locally.
The diagrams of the local spin current are shown in Fig. \ref{FD}.
The inhomogeneity of the external electric field is required when we discuss the leading effect driven by an inhomogeneous SOC strength.
Namely, the leading contribution of spin currents involving the derivative of the SOC strength has the form $\partial \lambda_{\mathrm{SO}} \partial \bm{E}_{\mathrm{em}}$.
The reason is as follows.
Spin current and electric field are odd under spatial inversion, while $\lambda_{\mathrm{SO}}$ and  other quantities are even with respect to spatial inversion. Therefore, the leading terms including spatial derivative of the SOC have the forms of $\partial^2 \lambda_{\mathrm{SO}} \bm{E}_{\mathrm{em}}$ or  $\partial \lambda_{\mathrm{SO}} \partial \bm{E}_{\mathrm{em}}$.  
We focus on the contributions $\partial \lambda_{\mathrm{SO}} \partial \bm{E}_{\mathrm{em}}$ by assuming $\partial^2 \lambda_{\mathrm{SO}}$ is negligibly small in this section. 
Now, we present the resulting local spin currents.
(For the detail of the calculation, refer to Appendix A.)
Under the condition $\epsilon_{F\sigma}\gg \hbar/2\tau_\sigma (\equiv \eta_\sigma)$ with $\tau_\sigma\equiv \hbar/2\pi u_0^2 n_{\mathrm{imp}} \nu_\sigma$ the spin-dependent relaxation time and $\nu_\sigma$ the density of state with spin index $\sigma=\pm1$, we obtain three types of the spin currents induced by an inhomogeneous SOC $(j_s^\alpha)^{\mathrm{local}}_i=(j_s^\alpha)^{(1)}_i +(j_s^\alpha)^{(2)}_i +(j_s^\alpha)^{(3)}_i$:
\begin{eqnarray}
(j_s^\alpha)^{(1)}_i &=& C (\alpha_1 \bm{e}_\alpha +\alpha_2 (\bm{e}_\alpha\times \bm{e}_z))\cdot (\nabla \lambda_{\mathrm{SO}}(\bm{r})\times \partial_i \bm{E}_{\mathrm{em}}(\bm{r})), \label{js1} \nonumber \\
\\
(j_s^\alpha)^{(2)}_i &=& C (\beta_1 (\bm{e}_i\times \bm{e}_\alpha) \nonumber \\
 & & +\beta_2 ( \delta_{iz} \bm{e}_\alpha -\delta_{i\alpha} \bm{e}_z)) \cdot \nabla \lambda_{\mathrm{SO}}(\bm{r}) (\nabla\cdot \bm{E}_{\mathrm{em}}(\bm{r})), \label{js2} \nonumber \\
 \\
(j_s^\alpha)^{(3)}_i &=& C (\gamma_1 \bm{e}_\alpha +\gamma_2 (\bm{e}_\alpha\times \bm{e}_z))\cdot (\nabla \lambda_{\mathrm{SO}}(\bm{r}) \times \nabla E^i_{\mathrm{em}}(\bm{r})), \label{js3} \nonumber \\
 & &
\end{eqnarray}
where $C \equiv \frac{\pi e \hbar^2}{2m} (\nu_\uparrow+\nu_\downarrow)$.
The detailed expressions of $\alpha_i, \beta_i,$ and $ \gamma_i\ (i=1,2)$ are given in Appendix A.
For $\alpha=z$, the second terms of Eqs. (\ref{js1}), (\ref{js2}) and (\ref{js3}) vanish.
It should be noted that even when $M$ goes to zero, the spin currents still have finite contributions and become isotropic in spin space.
Therefore, the magnetization is not indispensable for the spin-current generation itself.

\subsection{Diffusive spin current}
\begin{figure}[H]
 \begin{center} 
 \includegraphics[width=80mm]{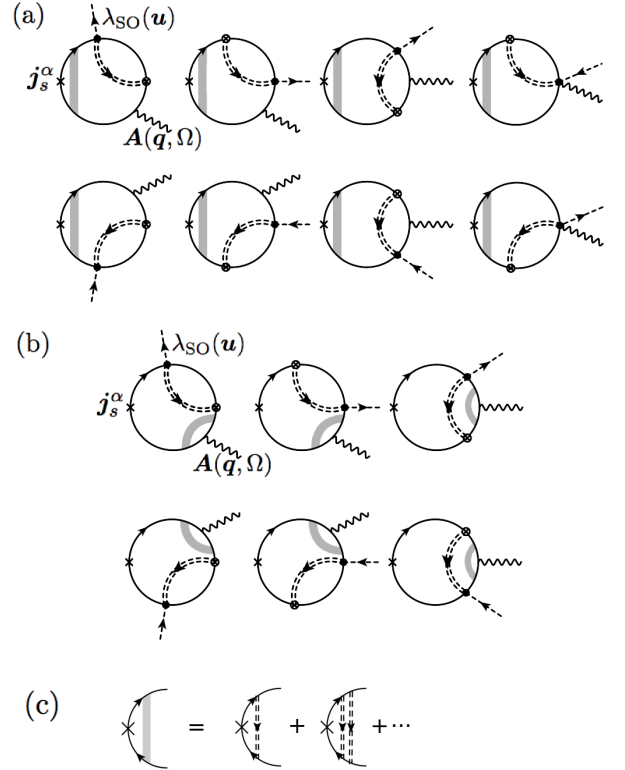} 
 \end{center} 
 \caption{Diagrams of the leading contribution of the diffusive spin current induced by the inhomogeneity of the SOC strength.
(a) Contributions with the vertex correction to the spin-current operator.
(b) Contributions with the vertex correction to the vector potential.
(c) The vertex correction is defined as the summation of the ladder diagram with respect to the normal impurity potential. 
} 
 \label{local} 
 \end{figure}
We calculate the diffusive (or non-local) spin current, which can be obtained by including the vertex correction to the local spin current.
Here, the vertex correction means the summation of the ladder diagram with respect to the normal impurity potential as shown in Fig. \ref{local} (c).
Since we focus on the diffusive behavior of the spin current induced by the exchange coupling, the ladder diagram does not involve the spin-orbit coupled impurities.
We consider two types of the vertex correction, i.e., that for the spin-current operator (Fig. \ref{local} (a)) and that for the vector potential (Fig. \ref{local} (b)).
The vertex correction that cuts across the impurity-averaged line is negligible because it is of higher order in $\eta_{\sigma}$ \cite{Hosono}.
For the diffusive contribution under the condition $\epsilon_{F\sigma}\gg \hbar/2\tau_\sigma$, we obtain two types of the spin currents induced by the inhomogeneous SOC $(j_s^\alpha)^{\mathrm{diffusive}}_i=(j_s^\alpha)^{(1')}_i+(j_s^\alpha)^{(2')}_i$:
\begin{widetext}
\begin{eqnarray}
(j_s^\alpha)^{(1')}_i &=& \left\{ \begin{array}{l}
 C \sum_\sigma ( \Re [\alpha_\sigma \bm{e}_\alpha \cdot \bm{K}^i_{(1),\sigma} (\bm{r})]+\Im [\alpha_\sigma (\bm{e}_\alpha\times \bm{e}_z) \cdot \bm{K}^i_{(2),\sigma} (\bm{r})]), \quad (\alpha=x,y) \label{jd} \\
 C \sum_\sigma \alpha_\sigma \bm{e}_z \cdot \bm{K}^i_{(2),\sigma} (\bm{r}), \quad (\alpha=z) \\
\end{array} \right. \\
\bm{K}^i_{(n),\sigma} (\bm{r}) &\equiv& \int d\bm{r}' \int d\bm{r}'' \int dt' \ \chi_\sigma^{(n)}(\bm{r}-\bm{r}',\bm{r}-\bm{r}'',t-t') ( \nabla \lambda_{\mathrm{SO}} (\bm{r}') \times \partial_i \bm{E}_{\mathrm{em}}(\bm{r}'',t') ), \ \ \ \ (n=1,2) \\
(j_s^\alpha)^{(2')}_i &=& C \sum_\sigma ( (\bm{e}_i\times \bm{e}_\alpha) \beta_{1, \sigma} +( \delta_{iz} \bm{e}_\alpha -\delta_{i\alpha} \bm{e}_z) \beta_{2, \sigma}) \cdot \nabla \lambda_{\mathrm{SO}} (\bm{r}) \int d\bm{r}' \int dt'\ \chi_\sigma (\bm{r}-\bm{r}',t-t') \nabla \cdot \bm{E}_{\mathrm{em}} (\bm{r}',t'). \label{je}
\end{eqnarray}
\end{widetext}
Here, 
\begin{eqnarray}
\chi^{(1)}_\sigma (\bm{r}_1,\bm{r}_2,t) &=& \int \frac{d\Omega}{2\pi} \sum_{\bm{u},\bm{q}} \frac{e^{i\Omega t+i\bm{u}\cdot \bm{r}_1+i\bm{q}\cdot \bm{r}_2}}{F_\sigma -i G_\sigma}, \label{chid} \\
\chi^{(2)}_\sigma (\bm{r}_1,\bm{r}_2,t) &=& \int \frac{d\Omega}{2\pi} \sum_{\bm{u},\bm{q}} \frac{e^{i\Omega t+i\bm{u}\cdot \bm{r}_1+i\bm{q}\cdot \bm{r}_2}}{(D_\sigma (\bm{u}+\bm{q})^2 -i\Omega) \tau_\sigma}, \\
\chi_\sigma (\bm{r},t) &=&  \int \frac{d\Omega}{2\pi} \sum_{\bm{q}} \frac{e^{i\Omega t+i\bm{q}\cdot \bm{r}}}{(D_\sigma \bm{q}^2 -i\Omega) \tau_\sigma},
\end{eqnarray}
represent the spin-diffusion propagators.
$D_\sigma\equiv \frac{2 \epsilon_F \tau_\sigma}{3m}$ denotes the diffusion coefficient.
The detailed expressions of $\alpha_\sigma, \beta_{i, \sigma}, F_\sigma,$ and $ G_\sigma\ (i=1,2)$ are given in Appendix B.
The vertex corrected spin current $(j_s^\alpha)^{(1')}_i$ ($(j_s^\alpha)^{(2')}_i$) corresponds to the local spin current $(j_s^\alpha)^{(1)}_i$ ($(j_s^\alpha)^{(2)}_i$).
There does not exist the diffusive-spin current contribution which corresponds to $(j_s^\alpha)^{(3)}_i$.
$(j_s^\alpha)^{(2')}_i$ differs from $(j_s^\alpha)^{(1')}_i$ in that only the divergence of the external field propagates.

\subsection{Spin-current absorption}

\begin{figure}
 \begin{center} 
 \includegraphics[width=80mm]{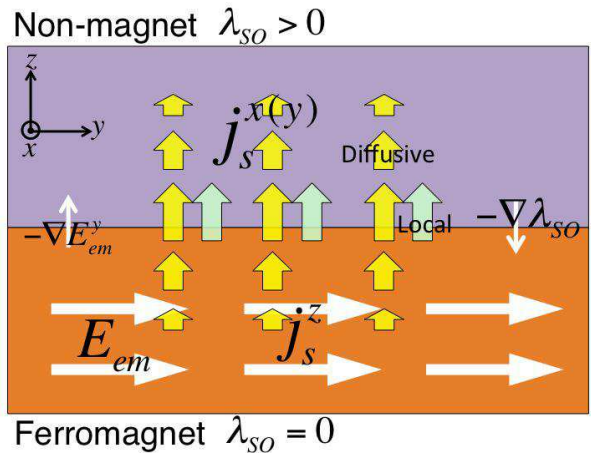} 
 \end{center} 
 \caption{(Color online) Schematic illustration of the spin-current absorption by the input spin-polarized current.
 The external electric field $\bm{E}_{\mathrm{em}}$ is applied to a ferromagnetic metal and therefore a spin-polarized current $\bm{j}_s^z$ flows in the direction of $y$ axis.
 The non-magnetic metal attached to the ferromagnetic metal has a strong SOC $\lambda_{\mathrm{SO}}$ due to impurities. 
 The diffusive spin currents $\bm{j}_s^{x,y}$ are induced perpendicularly to the interface and then decay exponentially in an oscillatory fashion.
 } 
 \label{sse} 
 \end{figure}

Let us consider a concrete configuration to show the absorption of the spin current by the inhomogeneous SOC.
We suppose that an external electric field is applied to a ferromagnetic metal and therefore a spin-polarized current flows in the $y$-direction  as shown in Fig. \ref{sse}.
Here, we assume that the applied electric field varies smoothly in space from the ferromagnet to the attached non-magnetic metal with SOC, which is modeled as $\bm{E}_{\mathrm{em}}(\bm{r})=E_y \frac{1+\tanh(-z/\xi_E)}{2}\bm{e}_y$ where $\xi_E$ is of the order of the lattice constant.
We also assume $\lambda_{\mathrm{SO}}(\bm{r})=\lambda_{\mathrm{SO}} \theta(z)$. In the perturbative calculation, we have considered a smoothly varying SOC strength. However,  this sharp spatial variation of the SOC would give a qualitatively correct results.\cite{Hosono}
In the present configuration, a spin-polarized current induced by the external electric field flows in the direction parallel to the interface.
Therefore, our setup is different from that of the spin-current injection.

The local spin current in the present configuration is generated  when $\lambda_{\mathrm{SO}}$ and $\bm{E}_{\mathrm{em}}$ vary in space, i.e., only near the interface between the ferromagnetic metal and the spin-orbit coupled metal.
Hence, the local spin current is not generated in the bulk spin-orbit coupled metal.
We thus focus on the non-local spin current as the spin-current absorption.
Since only $( \nabla \lambda_{\mathrm{SO}}(\bm{r}') \times \partial_i \bm{E}_{\mathrm{em}}(\bm{r}''))^j =E_y \lambda_{\mathrm{SO}} \delta(z') (2\xi_E \cosh^2(z''/\xi_E))^{-1} \delta_{iz} \delta_{j x}$ remains non-zero and $\nabla \cdot \bm{E}_{\mathrm{em}}=0$, the contribution of $(j_s^\alpha)^{(1')}_m$ (Eq.(11)) appears in the present configuration.
From Eq. (\ref{chid}), the spin-diffusive propagator in the static limit ($\Omega \to 0$) reduces to
\begin{eqnarray}
\chi^{(1)}_\sigma (\bm{r}_1,\bm{r}_2) &=& \frac{\pi}{2D_\sigma \tau_\sigma A_\sigma} \frac{e^{-\sqrt{B_\sigma/D_\sigma \tau_\sigma} |\bm{r}_1|}}{|\bm{r}_1|} \delta(\bm{r}_2-\bm{r}_1), \nonumber \\
\end{eqnarray}
where
\begin{eqnarray}
A_\sigma &\equiv& \frac{\eta_\sigma^2 \eta_+^2(\eta_+^2-3M^2)}{(M^2+\eta^2_+)^3}-i \frac{\sigma M \eta_\sigma^2 \eta_+ (3\eta_+^2-M^2)}{(M^2+\eta^2_+)^3}, \\
B_\sigma &\equiv& \frac{M(M^2+\eta_+^2)^2}{\eta_\sigma^2 \eta_+ (\eta_+^2 (\eta_+^2-3M^2)^2+M^2 (3\eta_+^2-M^2)^2)} \nonumber \\
 & & \times \left( 4M\eta_+(\eta_+^2-M^2) +i \sigma (6\eta_+^2M^2-M^4-\eta_+^4)\right), \nonumber \\
 & &
\end{eqnarray}
with $\eta_{\pm} \equiv \frac{\eta_\uparrow \pm \eta_\downarrow}{2}$.
Therefore, we obtain the spin currents for each spin component from Eq. (\ref{jd}):
\begin{eqnarray}
\bm{j}_s^x &=& \frac{\pi^2 C}{2\xi_E} E_y \lambda_{\mathrm{SO}} \sum_\sigma \frac{1}{D_\sigma \tau_\sigma} \Re \left[ \frac{\alpha_\sigma e^{-i|z| k^\sigma_s}}{A_\sigma \sqrt{B_\sigma}} \right] e^{-|z|/l^\sigma_s} \bm{e}_z, \label{31} \nonumber \\
\\
\bm{j}_s^y &=& \frac{\pi^2 C}{2\xi_E} E_y \lambda_{\mathrm{SO}} \sum_\sigma \frac{1}{D_\sigma \tau_\sigma} \Im \left[ \frac{\alpha_\sigma e^{-i|z| k^\sigma_s}}{A_\sigma \sqrt{B_\sigma}} \right] e^{-|z|/l^\sigma_s} \bm{e}_z, \label{32} \nonumber \\
\\
\bm{j}_s^z &=& \bm{0}.
\end{eqnarray}
Here
\begin{eqnarray}
l^\sigma_s &\equiv& \frac{\sqrt{D_\sigma \tau_\sigma}}{\left| \sqrt{B_\sigma} \right| \cos(\arg (\sqrt{B_\sigma}))}, \\
k^\sigma_s &\equiv& \sin (\arg (\sqrt{B_\sigma})),
\end{eqnarray}
which represent the spin diffusion length and the wave number for each spin, respectively. Here, we used the assumption $b\gg l_s^\sigma$ with $b$ the diameter of the interface \cite{Hosono}.

We find that spin currents are induced perpendicularly to the interface between the ferromagnetic metal and the spin-orbit coupled metal, and then decay exponentially in an oscillatory fashion as shown in Fig. \ref{sse}.
This exponential decay is due to the spin-flip scattering by the magnetization in the ferromagnet and also the oscillatory behavior originates from the difference between the Fermi wave numbers for each exchange spin-split bands.
The above spin currents are spin-polarized perpendicularly to the spin polarization of the input spin-polarized current (namely, $z$ component).
This reflects the violation of the conservation law of spin near the interface, and in other words, spin torque generated at the interface produces the present spin-current absorption.

It is also found that the magnitudes of the resulting spin currents in Eqs. (\ref{31}) and (\ref{32}) at $z=0$ are, respectively, proportional to $\eta_+^{1/2}$ and $\eta_- \eta_+^{-1/2}$ in the dirty limit ($\tau_\sigma \to 0\ (\sigma=\uparrow, \downarrow)$), and $\eta_- \eta_+^{1/2}$ and $\eta_+^{3/2}$ in the clean limit ($\tau_\sigma \to \infty\ (\sigma=\uparrow, \downarrow)$).
Namely, the magnitude of the absorbed spin currents gets reduced in clean ferromagnets.
This is due to the competition between the two effects of impurity scattering. 
In general, a SOC due to impurities in clean ferromagnets is weak due to low impurity concentration, leading to a small spin current. 
On the other hand, low impurity concentration leads to a large spin current since impurity scattering is suppressed.
When the former effect overcomes the latter, spin currents are expected to be suppressed in clean ferromagnets.
Finally, we note that for junctions composed of a weak spin-orbit coupled ferromagnet and a strong spin-orbit coupled non-magnet (namely, $\lambda_{\mathrm{SO}}(\bm{r})=\lambda^1_{\mathrm{SO}} \theta(z)+\lambda^2_{\mathrm{SO}} \theta(-z)$ with $\lambda^1_{\mathrm{SO}} > \lambda^2_{\mathrm{SO}}$), spin current absorption predicted in this paper also occurs.

\section{Absorption of spin current driven by spin accumulation}

\subsection{Model}

In this section, we examine the spin-current absorption originating from the variation of the SOC strength in systems with spin accumulation.
These systems are modeled by the conducting electrons in the presence of the gradient of the spin-dependent chemical potential and that of the SOC strength.
Our Hamiltonian consists of ${\cal H}_0, {\cal H}_{\mathrm{acc}}, {\cal H}_{\mathrm{imp}}$ and ${\cal H}^0_{\mathrm{SO}}$.
Here,
\begin{eqnarray}
{\cal H}_0 &=& \sum_{\sigma=\pm 1} \int d\bm{r}\ c_\sigma^\dagger(\bm{r},t)\left(-\frac{\hbar^2}{2m}\nabla^2 \right)c_\sigma(\bm{r},t), \\
{\cal H}_{\mathrm{acc}} &=& \sum_{\sigma=\pm 1} \int d\bm{r}\ \mu_\sigma(\bm{r},t)\ c_\sigma^\dagger(\bm{r},t)c_\sigma(\bm{r},t) \nonumber \\
 &=& \sum_{\sigma,\sigma'=\pm1} \int d\bm{r}\ c_\sigma^\dagger(\bm{r},t) (\bar{\mu}(\bm{r},t) \delta_{\sigma \sigma'}+\mu_s(\bm{r},t) \sigma^z_{{\sigma \sigma'}}) c_{\sigma'} (\bm{r},t). \nonumber \\
\end{eqnarray}
Here, we have defined the spin accumulation as the difference between spin-up and spin-down chemical potentials, namely, $\mu_s(\bm{r}):=\frac{\mu_\uparrow -\mu_\downarrow}{2}$. \cite{Kimura05, Takahashi03}
This spatial distribution of spin accumulation leads to a diffusive spin current. 
In the following, we will also assume the chemical potential $\bar{\mu}(\bm{r},t)$ to be constant.

\subsection{Spin-current}

We will calculate the spin current corresponding to the diagrams in Fig. \ref{local} in a similar manner to the previous section.
Here, we have replaced the vector potential $\bm{A}(\bm{q},\Omega)$ with the spin accumulation $\mu_s(\bm{q},\Omega)$ in the diagrams of Fig. \ref{local}.
We consider ${\cal H}_0$ as the non-perturbative part and ${\cal H}_{\mathrm{acc}}$, ${\cal H}_{\mathrm{imp}}$ and ${\cal H}^0_{\mathrm{SO}}$ as perturbative parts.
In this section, we adopt the self energy including both the normal and the spin-orbit coupled impurity potentials \cite{Hosono,Hikami}.
We also treat only the vertex correction for the spin-current operator \cite{Hosono}. This correction provides a finite decay length of the spin current determined by the SOC strength.
After taking the lesser component of the Green's functions and averaging over impurity positions, the spin current shown in Fig. \ref{FD} (a) and (b) reads
\begin{eqnarray}
(j_s^\alpha)_m &=& i \frac{\hbar^2}{m} \hbar \int \frac{d\Omega}{2\pi} \Omega e^{i\Omega t} \sum_{i,j,k}\sum_{\bm{k},\bm{k}',\bm{u},\bm{q}} \nonumber \\
 & & \times e^{i(\bm{u}+\bm{q})\cdot \bm{r}} \lambda_{\mathrm{SO}}(\bm{u}) \mu_s(\bm{q}, \Omega) \langle U(\bm{p})U(-\bm{p}) \rangle_{\mathrm{imp}} \nonumber \\
 & & \times \epsilon_{ijk} \Re (\chi_{\alpha ijkm}) \mathrm{tr}\left[ \sigma^\alpha \sigma^k \sigma^z \right].
\end{eqnarray}
For the expression of $\chi_{\alpha ijkm}$, refer to Appendix C.
In the same manner as in the previous section, we expand the coefficient $\chi_{\alpha ijkm}$ with respect to $\bm{u}$ and $\bm{q}$.
According to the inversion operation $\bm{r} \to -\bm{r}$, it is found that only the contributions with an odd order of $\bm{u}$ and $\bm{q}$ remain.
Now, we perform the vertex correction $\chi_{\alpha ijkm} \mathrm{tr}\left[ \sigma^\alpha \sigma^k \sigma^z \right] \to \tilde{\chi}_{\alpha ijk} \sum_{a,b,c,d,e} \sigma^\alpha_{da} \Gamma_{ad,cb} \sigma^k_{be} \sigma^z_{ec}$ with $\tilde{\chi}_{\alpha ijk}$ being the vertex-corrected coefficient and $\Gamma_{ad,cb}$ being the vertex correction corresponding to the ladder diagram including the normal and the spin orbit coupled impurities \cite{Hosono}.
The contributions from the vertex-corrected coefficient with the first order of $\bm{u}$ or $\bm{q}$ vanish, and hence the leading contribution in the diffusion regime is of the third order of $\bm{u} $ and $\bm{q}$:
$\tilde{\chi}_{\alpha ijk}=\tilde{\chi}^{(2,1)}_{\alpha ijk}+\tilde{\chi}^{(1,2)}_{\alpha ijk}+\cdots$, where the superscript $(i,j)$ denotes the order of $\bm{u}$ and $\bm{q}$.
The contributions from $\tilde{\chi}^{(3,0)}_{\alpha ijk}$ and $\tilde{\chi}^{(0,3)}_{\alpha ijk}$ vanish exactly.

From the contribution of $\tilde{\chi}^{(1,2)}_{\alpha ijkm}$, we obtain the expression of the spin current:
%
\begin{eqnarray}
(\tilde{j_s^\alpha})^{(1,2)}_m &=& \zeta_{(1,2)} \int d\bm{r}' \int d\bm{r}'' \int dt'\ D(\bm{r}-\bm{r}',\bm{r}-\bm{r}'',t-t') \nonumber \\
 & & \times \left[ \partial_\alpha \lambda_{\mathrm{SO}}(\bm{r}')\ \partial_z (\partial_m \mu_s(\bm{r}'', t')) \right. \nonumber \\
 & & \left. \ \ \  -\partial_z \lambda_{\mathrm{SO}}(\bm{r}')\ \partial_\alpha (\partial_m \mu_s(\bm{r}'',t')) \right],
\end{eqnarray}
%
where $ \zeta_{(1,2)}\equiv \frac{4\pi \hbar^2 \tau \nu}{3m}$ with $\tau\equiv \hbar/2\pi u_0^2 n_{\mathrm{imp}} \nu$ being the relaxation time and $\nu$ being the density of state.
%
\begin{eqnarray}
D(\bm{r}_1,\bm{r}_2,t) &:=& \int \frac{d\Omega}{2\pi} \sum_{\bm{u},\bm{q}} e^{i\Omega t+i\bm{u}\cdot \bm{r}_1+i\bm{q}\cdot \bm{r}_2} \nonumber \\
 & & \times \frac{\Omega (1-\kappa) (1+3\kappa)}{D\tau ((\bm{u}+\bm{q})^2+\xi^2) +i\Omega \tau}
\end{eqnarray}
is the spin-diffusive propagator with $\kappa\equiv \lambda_{\mathrm{SO}}^2 k_F^4/3$ ($k_F$ is the Fermi wave number), $D\equiv 2\epsilon_F \tau/3m$ and $\xi\equiv \sqrt{4\kappa/D\tau}$. As seen from Eq.(29),  $\xi$ determines the decay length of the spin current. 

As for the other contribution $\tilde{\chi}^{(2,1)}_{\alpha ijkm}$, we obtain the spin current of the form:
%
\begin{eqnarray}
(\tilde{j_s^\alpha})^{(2,1)}_m &=& \zeta_{(2,1)} \int d\bm{r}' \int d\bm{r}'' \int dt'\ D(\bm{r}-\bm{r}',\bm{r}-\bm{r}'',t-t') \nonumber \\
 & & \times \left[ \partial_\alpha \mu_s(\bm{r}'', t')\ \partial_z (\partial_m \lambda_{\mathrm{SO}}(\bm{r}')) \right. \nonumber \\
 & & \left. \ \ \  -\partial_z \mu_s(\bm{r}'', t')\ \partial_\alpha (\partial_m \lambda_{\mathrm{SO}}(\bm{r}')) \right],
\end{eqnarray}
%
where $\zeta_{(2,1)}\equiv \frac{608 \pi \hbar^2 \tau \nu}{45m}\left( \frac{\epsilon_F \tau}{\hbar} \right)^2$.
Here, we have focused on the diffusive spin currents, corresponding to contributions with the vertex correction.
It is found that if the scales of spatial variations of $\lambda_{\mathrm{SO}}(\bm{r})$ and $\mu_s(\bm{r},t)$ are nearly the same, the magnitude of the spin current $(\tilde{j_s^\alpha})^{(2,1)}_m$ is larger than that of $(\tilde{j_s^\alpha})^{(1,2)}_m$ by the factor of $\epsilon_F \tau/\hbar$.

\subsection{Spin-current absorption}

\begin{figure}
 \begin{center} 
 \includegraphics[width=95mm]{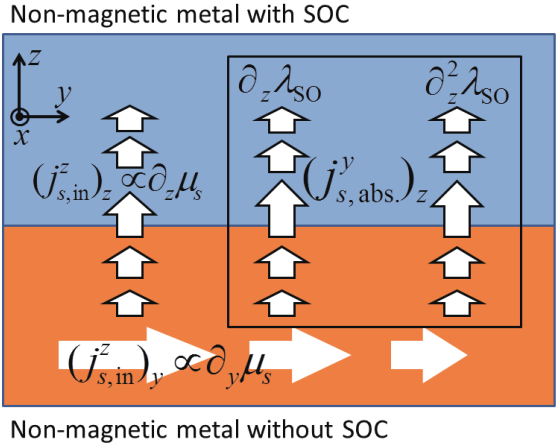} 
 \end{center} 
 \caption{(Color online) Schematic illustration of the absorption of spin currents driven by the spin accumulation.
 The arrows in the bottom metal denote input spin currents $(j^z_{s, \mathrm{in}})_y$ by the spin accumulation.
 In the top metal, the left arrows are normal diffusion of the input spin currents $(j^z_{s, \mathrm{in}})_z$, which is irrelevant to the variation of a SOC, whereas the middle and the right arrows are diffusion of spin currents $(j^y_{s, \mathrm{abs.}})_z$ induced in the vicinity of the interface, relevant to the difference of SOCs.
 } 
 \label{SA} 
 \end{figure}

Let us consider a typical configuration in order to investigate the spin-current absorption.
We assume that two non-magnetic metals with different SOC strengths are connected at the $z=0$ plane as shown in Fig. \ref{SA}.
In addition, there exists a gradient of the spin accumulation along the $y$ direction in one of the metals.
In the vicinity of the $z=0$ plane, we assume that $\lambda_{\mathrm{SO}}(\bm{r})=\lambda_{\mathrm{SO}}(z)$ and $\mu_s(\bm{r},t)=\mu_s(y,z,t)$, leading to
\begin{widetext}
\begin{eqnarray}
(\tilde{j_s^x})_z &=& 0, \\
(\tilde{j_s^y})_z &=& \int d\bm{r}' \int d\bm{r}'' \int dt'\ D(\bm{r}-\bm{r}',\bm{r}-\bm{r}'',t-t') \nonumber \\
 & & \times \left[ -\zeta_{(1,2)} \partial_{z''} \lambda_{\mathrm{SO}}(z'')\ \partial_{z'} (\partial_{y'} \mu_s(y',z',t')) +\zeta_{(2,1)} \partial_{y'} \mu_s(y',z',t')\ \partial^2_{z''} \lambda_{\mathrm{SO}}(z'') \right],\label{syz} \\ 
(\tilde{j_s^z})_z &=& 0.
\end{eqnarray}
\end{widetext}
%
%
%
The first term of the right hand side in Eq. (\ref{syz}) represents the interplay between the variation of the SOC strength and that of the input spin current near the interface. Note that spin current is induced by the gradient of the spin accumulation without SOC. \cite{Takahashi03}
On the other hand, the second term comes from the variation of the SOC strength and uniform input spin current.
Remark that a spin current polarized along the $y$-axis is generated as a response to that polarized along the $z$-axis.
Similar to the absorption of spin currents driven by an external electric field, a spin current absorbed into the top metal is non-local, and diffusive spin current decays due to the spin-orbit coupled scattering as seen from Eq.(29).

\section{Discussion}

In the present study, we have focused on diffusion of spin currents including the gradient of the SOC strength.
However, there also exists a diffusion of spin currents irrelevant to the difference of SOC strength as shown in Fig. \ref{SA}. 
We can extract the former contribution by comparing spin currents in the triple lateral spin valves using middle junctions with different SOC \cite{Kimura08}.
In fact, the absorption of non-local spin currents has been successfully demonstrated in the spin-valve measurement consisting of double Py/Cu junctions and middle Au/Cu junction \cite{Kimura07, Kimura08}. 

Since the argument based on the mismatch of the spin resistance by Kimura \textit{et al.} \cite{Kimura05} assumes the continuity of a spin current, the absorbed spin currents are not those generated at an interface.
Consequently, the absorbed spin currents in our theory are of different origin from those in the theory based on the spin-resistance mismatch.
Hence, we cannot compare our results with previous results directly.
Both studies show the spin-current absorption with different mechanisms.

In asymmetric structures, conduction electrons at its interface in general feel a Rashba SOC \cite{Rashba60}, leading to the spin polarization under an external electric field or current injection \cite{Edelstein90}.
In the configurations of Figs. \ref{sse} and \ref{SA}, there would exist a Rashba SOC at their interfaces.
The current-induced spin polarization due to the Rashba SOC may affect the absorption of spin currents predicted in our models. 


\section{Summary}

We have investigated the generation of spin currents by an inhomogeneous SOC due to impurities, which have been applied to the interface system to show the spin-current absorption.
Using the Keldysh Green's function formalism, we have presented analytical expressions of the spin currents with the gradient of the SOC strength for two systems: the system with field-driven spin currents and one with spin-accumulation-driven spin currents.
The resulting spin current indicates the absorption of the spin current at the interface between materials with different SOC strengths.

In the present study, we assumed a homogeneous magnetization.
The extension of our model to an inhomogeneous magnetization is also an interesting future work. 

\begin{acknowledgments}
We thank Y. Tserkovnyak, G. Tatara and A. Takeuchi for helpful discussions.
K. T. thanks S. Murakami for discussions.
K. H. thanks Y. Nozaki for discussions.
This work is supported by Grant-in-Aid for Young Scientists (B) (No. 23740236 and No. 24710153) and the "Topological Quantum Phenomena" (No. 23103505) Grant-in Aid for Scientific Research on Innovative Areas from the Ministry of Education, Culture, Sports, Science and Technology (MEXT) of Japan.
K. T. also acknowledges the financial support from the Global Center of Excellence Program by MEXT, Japan, through the gNanoscience and Quantum Physicsh Project of the Tokyo Institute of Technology.
\end{acknowledgments}

\appendix

\onecolumngrid

\section{Calculation of local spin current}

We perturbatively calculate the spin currents induced by the inhomogeneity of the SOC strength using the Keldysh Green's function formalism.
We first treat the local spin current, which is locally driven by the external electric field.
The leading contribution of the spin current involves the first order with respect to the SOC strength, which is diagrammatically shown in Fig. \ref{FD}. The local spin currents with $\alpha$-component spin polarization flowing in the $m$ direction, i.e., $(j_s^\alpha)^{\mathrm{local}}_m=(j_s^\alpha)^{\mathrm{sj}}_m+(j_s^\alpha)^{\mathrm{sk}}_m$ thus  read
\begin{eqnarray}
(j_s^\alpha)^{\mathrm{sj}}_m &=& i \frac{\hbar^2}{2m} \hbar \frac{e\hbar}{2m} \int \frac{d\Omega}{2\pi} \int \frac{d\omega}{2\pi} \Omega e^{i\Omega t} \sum_{i,j,k,l}\sum_{\bm{k},\bm{k}',\bm{u},\bm{q}} e^{i(\bm{u}+\bm{q})\cdot \bm{r}} \lambda_{\mathrm{SO}}(\bm{u}) \langle U(\bm{p})U(-\bm{p}) \rangle_{\mathrm{imp}} A_l(\bm{q},\Omega) \epsilon_{ijk}\ 2\Re (\chi_{\alpha ijklm}), \label{jsj} \\
(j_s^\alpha)^{\mathrm{sk}}_m &=& i \frac{\hbar^2}{2m} e \int \frac{d\Omega}{2\pi} \int \frac{d\omega}{2\pi} \Omega e^{i\Omega t} \sum_{i,j,k}\sum_{\bm{k},\bm{k}',\bm{u},\bm{q}} e^{i(\bm{u}+\bm{q})\cdot \bm{r}} \lambda_{\mathrm{SO}}(\bm{u}) \langle U(\bm{p})U(-\bm{p}) \rangle_{\mathrm{imp}} A_j(\bm{q},\Omega) \epsilon_{ijk}\ 2\Re(\chi_{\alpha ikm}) \label{jsk},
\end{eqnarray}
where the superscripts "sj" and "sk" denote the side-jump and the skew-scattering contributions, respectively,\cite{Nagaosa} and each coefficients are given by
\begin{eqnarray}
\chi_{\alpha ijklm} &=& +(k-u-k')_i (k+k')_j (2k-2u-q)_l (2k-u-q)_m \mathrm{tr}\left[ \sigma^\alpha g^R_{\bm{k}} \sigma^k g^R_{\bm{k}'} g^R_{\bm{k}-\bm{u}} g^A_{\bm{k}-\bm{u}-\bm{q}} \right] \nonumber \\
 & & +(k'-k)_i (k-u+k')_j (2k-2u-q)_l (2k-u-q)_m \mathrm{tr}\left[ \sigma^\alpha g^R_{\bm{k}} g^R_{\bm{k}'} \sigma^k  g^R_{\bm{k}-\bm{u}} g^A_{\bm{k}-\bm{u}-\bm{q}} \right] \nonumber \\
 & & +(k-u-k')_i (k+k')_j (2k'-q)_l (2k-u-q)_m \mathrm{tr}\left[ \sigma^\alpha g^R_{\bm{k}} \sigma^k g^R_{\bm{k}'} g^A_{\bm{k}'-\bm{q}} g^A_{\bm{k}-\bm{u}-\bm{q}} \right], \\
 \chi_{\alpha ikm} &=& (k-k'-u-q)_i (2k-u-q)_m \mathrm{tr}\left[ \sigma^\alpha g^R_{\bm{k}} \sigma^k g^A_{\bm{k}'} g^A_{\bm{k}-\bm{u}-\bm{q}} \right].
\end{eqnarray}
Here, $g^{R(A)}_{\bm{k}}\equiv g^{R(A)}_{\bm{k},\omega=0}$, and $g^R_{\bm{k},\omega}$ and $g^A_{\bm{k},\omega}$
denote the non-perturbative retarded and advanced Green's function, respectively, which are $2 \times 2$ matrices in spin space.
We also use the Langreth's method to obtain $[g_{\bm{k}_1,\omega} \cdots g_{\bm{k}_i,\omega} g_{\bm{k}_{i+1},\omega-\Omega} \cdots g_{\bm{k}_n,\omega-\Omega}]^< \simeq -\Omega f'(\omega) g_{\bm{k}_1,\omega}^R \cdots g_{\bm{k}_i,\omega}^R g_{\bm{k}_{i+1},\omega}^A \cdots g_{\bm{k}_n,\omega}^A$ with $f(\omega)$ the Fermi distribution function \cite{Hung98}.
Here, we set temperature to zero, i.e., $f'(\omega)\simeq -\delta(\omega)$.

Since we are interested in spin currents produced by the inhomogeneous SOC strength, we focus on the contribution which involves the spatial derivative of the SOC strength.
After some algebraic calculations, we obtain the above coefficients represented as 
\begin{eqnarray}
\chi_{\alpha ijklm} &\simeq& \frac{2}{3}k^2 (\delta_{jl} u_i q_m +\delta_{jm} u_i q_l) \left( \mathrm{tr}\left[ \sigma^\alpha g^R_{\bm{k}} \sigma^k g^R_{\bm{k}'} g^R_{\bm{k}} g^A_{\bm{k}} \right] +\mathrm{tr}\left[ \sigma^\alpha g^R_{\bm{k}} g^R_{\bm{k}'} \sigma^k  g^R_{\bm{k}} g^A_{\bm{k}} \right] \right) \nonumber \\
 & & +\frac{4}{15} \frac{\hbar^2}{m} k^4 (\delta_{jl} u_i q_m+\delta_{jm} u_i q_l+\delta_{lm} u_i q_j) \left( \mathrm{tr}\left[ \sigma^\alpha g^R_{\bm{k}} \sigma^k g^R_{\bm{k}'} g^R_{\bm{k}} (g^A_{\bm{k}})^2 \right] +\mathrm{tr}\left[ \sigma^\alpha g^R_{\bm{k}} g^R_{\bm{k}'} \sigma^k  g^R_{\bm{k}} (g^A_{\bm{k}})^2 \right] \right) \nonumber \\
 & & +\frac{2}{3} \left( k'^2 \delta_{jl} u_i q_m +k^2 \delta_{jm} u_i q_l \right) \mathrm{tr}\left[ \sigma^\alpha g^R_{\bm{k}} \sigma^k g^R_{\bm{k}'} g^A_{\bm{k}'} g^A_{\bm{k}} \right] \nonumber \\
 & & +\frac{8}{9} \frac{\hbar^2}{m} k^2k'^2 \delta_{jl} u_i q_m \mathrm{tr}\left[ \sigma^\alpha g^R_{\bm{k}} \sigma^k g^R_{\bm{k}'} g^A_{\bm{k}'} (g^A_{\bm{k}})^2 \right] +\frac{4}{9} \frac{\hbar^2}{m} k^2k'^2 \delta_{jm} u_i q_l \mathrm{tr}\left[ \sigma^\alpha g^R_{\bm{k}} \sigma^k g^R_{\bm{k}'} (g^A_{\bm{k}'})^2 g^A_{\bm{k}} \right], \label{chi1} \\
 \chi_{\alpha ikm} &\simeq& (u_i q_m+q_i u_m) \left( \mathrm{tr}\left[ \sigma^\alpha g^R_{\bm{k}} \sigma^k g^A_{\bm{k}'} g^A_{\bm{k}} \right] + \frac{\hbar^2}{m} k^2 \mathrm{tr}\left[ \sigma^\alpha g^R_{\bm{k}} \sigma^k g^A_{\bm{k}'} (g^A_{\bm{k}})^2 \right] \right). \label{chi2}
\end{eqnarray}
Here, we assumed the rotational symmetry of the system, i.e., $\sum_{\bm{k}} k_i k_j=\sum_{k} \frac{k^2}{3} \delta_{ij}$ and $\sum_{\bm{k}} k_i k_j k_l k_m=\sum_{k} \frac{k^4}{15} (\delta_{ij}\delta_{lm}+\delta_{il}\delta_{jm}+\delta_{im}\delta_{jl})$.
We remark that if we do not consider the inhomogeneity of the external electric field, the present contributions vanish.
The inhomogeneity of the external electric field is required to obtain the contributions from the inhomogeneous SOC strength.

Next, we take traces over spin space in Eqs. (\ref{chi1}) and (\ref{chi2}) by using the formula $\mathrm{tr}[\sigma^\alpha A \sigma^k B]=\sum_{\sigma=\pm 1} (\delta_{\alpha k}-\delta_{\alpha z}\delta_{kz}-i\sigma \epsilon_{\alpha z k}) A_{\sigma} B_{\bar{\sigma}} +\sum_{\sigma=\pm 1} \delta_{\alpha z}\delta_{kz} A_\sigma B_\sigma$ with $A, B$ Green's functions, and then Eqs. (\ref{jsj}) and (\ref{jsk}) reduce to
\begin{eqnarray}
(j_s^\alpha)^{\mathrm{sj}}_m &\simeq& \frac{e \hbar^2}{2m} u^2_0 n_{\mathrm{imp}} \int \frac{d\Omega}{2\pi} \ \Omega e^{i\Omega t} \sum_{\bm{u},\bm{q}} \sum_{i,j,k,l} e^{i(\bm{u}+\bm{q})\cdot \bm{r}} \lambda_{\mathrm{SO}}(\bm{u}) A_l(\bm{q},\Omega) \epsilon_{ijk} \sum_\sigma ((\delta_{\alpha k}-\delta_{\alpha z}\delta_{kz}) \Re +\sigma \epsilon_{\alpha k} \Im) \nonumber \\
 & & \times \left[ \frac{2}{3} \delta_{jl} u_i q_m \left( (I^\sigma_{01;00}+ I^\sigma_{10;00}) \left( Q^\sigma_{11;01}+\frac{4}{5} S^\sigma_{11;02} \right) +Q^\sigma_{01;01} I^\sigma_{10;01} +\frac{8}{3} Q^\sigma_{01;01} Q^\sigma_{10;02} \right) \right. \nonumber \\
 & & +\frac{2}{3} \delta_{jm} u_i q_l \left( \left( I^\sigma_{01;00}+ I^\sigma_{10;00} \right) \left( Q^\sigma_{11;01}+\frac{4}{5} S^\sigma_{11;02} \right) +I^\sigma_{01;01} Q^\sigma_{10;01} +\frac{4}{3} Q^\sigma_{01;02} Q^\sigma_{10;01} \right) \nonumber \\
 & & \left. + \frac{8}{15} \delta_{lm} u_i q_j (I^\sigma_{01;00}+ I^\sigma_{10;00}) S^\sigma_{11;02} \right] \nonumber \\
 & & +\frac{e \hbar^2}{2m} u^2_0 n_{\mathrm{imp}} \int \frac{d\Omega}{2\pi} \ \Omega e^{i\Omega t} \sum_{\bm{u},\bm{q}} \sum_{i,j,k,l} e^{i(\bm{u}+\bm{q})\cdot \bm{r}} \lambda_{\mathrm{SO}}(\bm{u}) A_l(\bm{q},\Omega) \epsilon_{ijk} \sum_\sigma \delta_{\alpha z}\delta_{kz} \Re \nonumber \\
 & & \times \left[ \frac{2}{3} \delta_{jl} u_i q_m \left( 2I^\sigma_{10;00} \left( Q^\sigma_{20;10}+\frac{4}{5} S^\sigma_{20;20} \right) +Q^\sigma_{10;10} I^\sigma_{10;10} +\frac{8}{3} Q^\sigma_{10;10} Q^\sigma_{10;20} \right) \right. \nonumber \\
 & & +\frac{2}{3} \delta_{jm} u_i q_l \left( 2I^\sigma_{10;00} \left( Q^\sigma_{20;10}+\frac{4}{5} S^\sigma_{20;20} \right) +I^\sigma_{10;10} Q^\sigma_{10;10} +\frac{4}{3} Q^\sigma_{10;20} Q^\sigma_{10;10} \right) \nonumber \\
 & & \left. + \frac{16}{15} \delta_{lm} u_i q_j I^\sigma_{10;00} S^\sigma_{20;20} \right], \label{sdjp} \\
(j_s^\alpha)^{\mathrm{sk}}_m &\simeq& \frac{e \hbar^2}{2m} u^2_0 n_{\mathrm{imp}} \int \frac{d\Omega}{2\pi} \ \Omega e^{i\Omega t} \sum_{i,j,k} \sum_{\bm{u},\bm{q}} e^{i(\bm{u}+\bm{q})\cdot \bm{r}} \lambda_{\mathrm{SO}}(\bm{u})A_j(\bm{q},\Omega) \epsilon_{ijk} \left[ \sum_\sigma (\delta_{\alpha k}-\delta_{\alpha z}\delta_{kz}) \Re +\sigma \epsilon_{\alpha k} \Im) \right. \nonumber \\
 & & \left. \times u_i q_m I^\sigma_{00;01} (I^\sigma_{10;01} + 2Q^\sigma_{10;02} ) +\sum_\sigma \delta_{\alpha z}\delta_{kz} u_i q_m \Re I^\sigma_{00;10} (I^\sigma_{10;10} + 2Q^\sigma_{10;20} ) \right] \label{skst}.
\end{eqnarray}
Here, we define integrals of the Green's functions appearing in the above expressions as
\begin{eqnarray}
I^\sigma_{ab;cd} &\equiv& \sum_{\bm{k}} (g^R_{\bm{k},\sigma})^a (g^R_{\bm{k},\bar{\sigma}})^b (g^A_{\bm{k},\sigma})^c (g^A_{\bm{k},\bar{\sigma}})^d, \\
Q^\sigma_{ab;cd} &\equiv& \sum_{\bm{k}} \epsilon_{\bm{k}} (g^R_{\bm{k},\sigma})^a (g^R_{\bm{k},\bar{\sigma}})^b (g^A_{\bm{k},\sigma})^c (g^A_{\bm{k},\bar{\sigma}})^d, \\
S^\sigma_{ab;cd} &\equiv& \sum_{\bm{k}} \epsilon^2_{\bm{k}} (g^R_{\bm{k},\sigma})^a (g^R_{\bm{k},\bar{\sigma}})^b (g^A_{\bm{k},\sigma})^c (g^A_{\bm{k},\bar{\sigma}})^d,
\end{eqnarray}
where $g^R_{\bm{k},\sigma}=[\epsilon_{F\sigma}-\epsilon_{\bm{k}}+i\eta_\sigma]^{-1}$ with $\eta_\sigma \equiv \hbar/2\tau_\sigma$ and $g^A_{\bm{k},\sigma}=(g^R_{\bm{k},\sigma})^*$ with $\epsilon_{\bm{k}}\equiv \frac{\hbar^2 k^2}{2m}$.
We sum up Eqs. (\ref{sdjp}) and (\ref{skst}) and consider the dominant contribution from the integrals of the Green's functions under the condition $\epsilon_{F\sigma}\gg \hbar/2\tau_\sigma$.
By carrying out the integrals of the Green's functions in Eqs.(A9-A11) and transforming to the real-space representation, we obtain the local spin current of the form
\begin{eqnarray}
(j_s^\alpha)_m &\simeq& \frac{\pi e \hbar^2}{2m} (\nu_\uparrow+\nu_\downarrow) \left[ (\alpha_1 \bm{e}_\alpha +\alpha_2 (\bm{e}_\alpha\times \bm{e}_z))\cdot (\nabla \lambda_{\mathrm{SO}}(\bm{r})\times \partial_m \bm{E}_{\mathrm{em}}(\bm{r})) \right. \nonumber \\
 & & \ \ \ \ \ \ \ \ +(\beta_1 (\bm{e}_m\times \bm{e}_\alpha) +\beta_2 ( \delta_{mz} \bm{e}_\alpha -\delta_{m\alpha} \bm{e}_z)) \cdot \nabla \lambda_{\mathrm{SO}}(\bm{r}) (\nabla\cdot \bm{E}_{\mathrm{em}}(\bm{r})) \nonumber \\
 & & \ \ \ \ \ \ \ \ \left. +(\gamma_1 \bm{e}_\alpha +\gamma_2 (\bm{e}_\alpha\times \bm{e}_z))\cdot (\nabla \lambda_{\mathrm{SO}}(\bm{r}) \times \nabla E^m_{\mathrm{em}}(\bm{r})) \right].
\end{eqnarray}
Here, each coefficients are  dimensionless and given by
\begin{eqnarray}
\alpha_i &\equiv& \gamma_i+\delta_i, \\
\beta_i &\equiv& \gamma_i+\epsilon_i, \\
\delta_1 &\equiv& \left\{ \begin{array}{l}
\frac{16}{9} \left( \frac{\epsilon_F M^2 \eta_+}{(M^2+\eta^2_+)^2} -\frac{1}{4}\frac{1}{\nu_\uparrow+\nu_\downarrow}\frac{\eta_+}{M^2+\eta^2_+} \sum_\sigma \epsilon_{F\sigma}\nu_{\bar{\sigma}} \right), \quad (\alpha=x,y) \\
-\frac{4}{9} \frac{\epsilon_F}{\eta_+}, \quad (\alpha=z) \\
\end{array} \right. \\
\delta_2 &\equiv& \frac{16}{9} \left( \frac{\epsilon_F M(M^2-\eta^2_+)}{(M^2+\eta^2_+)^2} -\frac{1}{4}\frac{1}{\nu_\uparrow+\nu_\downarrow}\frac{M}{M^2+\eta^2_+} \sum_\sigma \epsilon_{F\sigma}\nu_{\bar{\sigma}} \right), \\
\epsilon_1 &\equiv& \left\{ \begin{array}{l}
\frac{8}{9} \left( \frac{M}{M^2+\eta^2_+} \sum_\sigma \sigma \epsilon_{F\sigma} \frac{\epsilon_{F\sigma}}{\eta_\sigma} +\frac{2}{\nu_\uparrow+\nu_\downarrow} \sum_\sigma \frac{\epsilon_{F\sigma}}{\eta_\sigma} \nu_{\bar{\sigma}} \right), \quad (\alpha=x,y) \\
\frac{16}{9} \frac{\epsilon_{F\sigma}}{\eta_\sigma}, \quad (\alpha=z) \\
\end{array} \right. \\
\epsilon_2 &\equiv& \frac{8}{9} \frac{\eta_+}{M^2+\eta^2_+} \sum_\sigma \sigma \epsilon_{F\sigma} \frac{\epsilon_{F\sigma}}{\eta_\sigma}, \\
\gamma_1 &\equiv& \left\{ \begin{array}{l}
\frac{2}{15} \left[ \frac{M \eta_+}{M^2+\eta^2_+} \sum_\sigma \sigma \left( \frac{\epsilon_{F\sigma}}{\eta_\sigma}\right)^2 +\frac{M \eta_-}{M^2+\eta^2_-} \sum_\sigma \left( \frac{\epsilon_{F\sigma}}{\eta_\sigma}\right)^2 -\frac{M \eta^2_+}{(M^2+\eta^2_+)^2} \sum_\sigma \sigma \epsilon_{F\sigma} \frac{\epsilon_{F\sigma}}{\eta_\sigma} -\frac{\epsilon_F \eta_+}{M^2+\eta^2_+} \right], \quad (\alpha=x,y) \\
-\frac{2}{15} \frac{\epsilon_F}{\eta_+}, \quad (\alpha=z) \\
\end{array} \right. \\
\gamma_2 &\equiv& -\frac{2}{15} \left[ \frac{M^2(\eta^2_+ -\eta_-^2)}{(M^2+\eta^2_+)(M^2+\eta^2_-)} \sum_\sigma \sigma \left( \frac{\epsilon_{F\sigma}}{\eta_\sigma}\right)^2 +\frac{\eta_+ (M^2-\eta^2_+)}{(M^2+\eta^2_+)^2} \sum_\sigma \sigma \epsilon_{F\sigma} \frac{\epsilon_{F\sigma}}{\eta_\sigma} +\frac{\epsilon_F M}{M^2+\eta^2_+} \right],
\end{eqnarray}
and $\eta_{\pm} \equiv \frac{\eta_\uparrow \pm \eta_\downarrow}{2}$.

\section{Calculation of diffusive spin current}

We calculate the diffusive (or non-local) spin current, which is represented by the vertex correction to the local spin current.
This can be performed by replacing $\chi_{\alpha ijklm}$ and $ \chi_{\alpha ikm}$ in Eqs. (\ref{jsj}) and (\ref{jsk}) with $\tilde{\chi}_{\alpha ijklm}=\Gamma^m_{\bm{u},\bm{q}} \Pi_{\bm{u},\bm{q}} \chi^{(1)}_{\alpha ijkl} +\Gamma^l_{\bm{q}} \Pi_{\bm{q}} \chi^{(2)}_{\alpha ijkm}$ and $\ \tilde{\chi}_{\alpha ikm}=\Gamma^m_{\bm{u},\bm{q}} \Pi_{\bm{u},\bm{q}} \chi_{\alpha ik}$, respectively, where
\begin{eqnarray}
\Gamma^m_{\sigma, \bar{\sigma}} (\bm{u}) &=& \frac{u^2_0 n_{\mathrm{imp}}}{V} \sum_{\bm{k}} (2k-u)_m g^R_{\bm{k},\omega,\sigma} g^A_{\bm{k}-\bm{u},\omega-\Omega,\bar{\sigma}}, \\
\Pi_{\sigma, \bar{\sigma}} (\bm{u}) &=& \sum_{n=0}^\infty \left( \frac{u^2_0 n_{\mathrm{imp}}}{V} \sum_{\bm{k}} g^R_{\bm{k},\omega,\sigma} g^A_{\bm{k}-\bm{u},\omega-\Omega,\bar{\sigma}} \right)^n, 
\\
\\
\chi^{(1)}_{\alpha ijkl} &=& +(k-u-k')_i (k+k')_j (2k-2u-q)_l \mathrm{tr}\left[ \sigma^\alpha g^R_{\bm{k}} \sigma^k g^R_{\bm{k}'} g^R_{\bm{k}-\bm{u}} g^A_{\bm{k}-\bm{u}-\bm{q}} \right] \nonumber \\
 & & +(k'-k)_i (k-u+k')_j (2k-2u-q)_l \mathrm{tr}\left[ \sigma^\alpha g^R_{\bm{k}} g^R_{\bm{k}'} \sigma^k  g^R_{\bm{k}-\bm{u}} g^A_{\bm{k}-\bm{u}-\bm{q}} \right] \nonumber \\
 & & +(k-u-k')_i (k+k')_j (2k'-q)_l \mathrm{tr}\left[ \sigma^\alpha g^R_{\bm{k}} \sigma^k g^R_{\bm{k}'} g^A_{\bm{k}'-\bm{q}} g^A_{\bm{k}-\bm{u}-\bm{q}} \right], \\
\chi^{(2)}_{\alpha ijkm} &=& +(k-u-k')_i (k+k')_j (2k-u-q)_m \mathrm{tr}\left[ \sigma^\alpha g^R_{\bm{k}} \sigma^k g^R_{\bm{k}'} g^R_{\bm{k}-\bm{u}} g^A_{\bm{k}-\bm{u}-\bm{q}} \right] \nonumber \\
 & & +(k'-k)_i (k-u+k')_j (2k-u-q)_m \mathrm{tr}\left[ \sigma^\alpha g^R_{\bm{k}} g^R_{\bm{k}'} \sigma^k  g^R_{\bm{k}-\bm{u}} g^A_{\bm{k}-\bm{u}-\bm{q}} \right] \nonumber \\
 & & +(k-u-k')_i (k+k')_j (2k-u-q)_m \mathrm{tr}\left[ \sigma^\alpha g^R_{\bm{k}} \sigma^k g^R_{\bm{k}'} g^A_{\bm{k}'-\bm{q}} g^A_{\bm{k}-\bm{u}-\bm{q}} \right], \\
 \chi_{\alpha ik} &=& (k-k'-u-q)_i \mathrm{tr}\left[ \sigma^\alpha g^R_{\bm{k}} \sigma^k g^A_{\bm{k}'} g^A_{\bm{k}-\bm{u}-\bm{q}} \right].
\end{eqnarray}
Expanding with respect to $\bm{u}$ or $\bm{q}$, we obtain the leading contributions as follows
\begin{eqnarray}
\chi^{(1)}_{\alpha ijkl} &\simeq& -\frac{2}{3} k^2 \delta_{jl} u_i \left( \mathrm{tr}\left[ \sigma^\alpha g^R_{\bm{k}} \sigma^k g^R_{\bm{k}'} g^R_{\bm{k}} g^A_{\bm{k}} \right] +\mathrm{tr}\left[ \sigma^\alpha g^R_{\bm{k}} g^R_{\bm{k}'} \sigma^k  g^R_{\bm{k}} g^A_{\bm{k}} \right] \right) -\frac{2}{3} k'^2 \delta_{jl} u_i \mathrm{tr}\left[ \sigma^\alpha g^R_{\bm{k}} \sigma^k g^R_{\bm{k}'} g^A_{\bm{k}'} g^A_{\bm{k}} \right] \nonumber \\
 & & -\frac{4}{9} \frac{\hbar^2}{m} k^2k'^2 \delta_{jl} (u+q)_i \mathrm{tr}\left[ \sigma^\alpha g^R_{\bm{k}} \sigma^k g^R_{\bm{k}'} g^A_{\bm{k}'} (g^A_{\bm{k}})^2 \right], \label{ch4} \\
\chi^{(2)}_{\alpha ijkm} &\simeq& -\frac{2}{3} k^2 \delta_{jm} u_i \left( \mathrm{tr}\left[ \sigma^\alpha g^R_{\bm{k}} \sigma^k g^R_{\bm{k}'} g^R_{\bm{k}} g^A_{\bm{k}} \right] +\mathrm{tr}\left[ \sigma^\alpha g^R_{\bm{k}} g^R_{\bm{k}'} \sigma^k  g^R_{\bm{k}} g^A_{\bm{k}} \right] \right) -\frac{2}{3} k^2 \delta_{jm} u_i \mathrm{tr}\left[ \sigma^\alpha g^R_{\bm{k}} \sigma^k g^R_{\bm{k}'} g^A_{\bm{k}'} g^A_{\bm{k}} \right] \nonumber \\
 & & -\frac{4}{9} \frac{\hbar^2}{m} k^2k'^2 \delta_{im} q_j \mathrm{tr}\left[ \sigma^\alpha g^R_{\bm{k}} \sigma^k g^R_{\bm{k}'} (g^A_{\bm{k}'})^2 g^A_{\bm{k}} \right], \label{ch5} \\
\chi_{\alpha ik} &\simeq& -(u+q)_i \left( \mathrm{tr}\left[ \sigma^\alpha g^R_{\bm{k}} \sigma^k g^A_{\bm{k}'} g^A_{\bm{k}} \right] +\frac{1}{3}\frac{\hbar^2}{m} k^2 \mathrm{tr}\left[ \sigma^\alpha g^R_{\bm{k}} \sigma^k g^A_{\bm{k}'} (g^A_{\bm{k}})^2 \right] \right). \label{ch3}
\end{eqnarray}
Next, we take traces over spin space in Eqs. (\ref{ch4}), (\ref{ch5}) and (\ref{ch3}) in a similar manner to the local spin current, and then the diffusive spin currents are reduced to
\begin{eqnarray}
(j_s^\alpha)^{\mathrm{sj}}_m &\simeq& \frac{e \hbar^2}{2m} u^2_0 n_{\mathrm{imp}} \int \frac{d\Omega}{2\pi} \ \Omega e^{i\Omega t} \sum_{\bm{u},\bm{q}} \sum_{i,j,k,l} e^{i(\bm{u}+\bm{q})\cdot \bm{r}} \lambda_{\mathrm{SO}}(\bm{u}) A_l(\bm{q},\Omega) \epsilon_{ijk} \sum_\sigma ((\delta_{\alpha k}-\delta_{\alpha z}\delta_{kz}) \Re +\sigma \epsilon_{\alpha k} \Im) \nonumber \\
 & & \times \left[ \frac{2}{3} \delta_{jl} u_i q_m\ \Pi_{\sigma, \bar{\sigma}} (\bm{u}+\bm{q}) u_0^2 n_{\mathrm{imp}} I^\sigma_{10;01} \left( (I^\sigma_{01;00}+ I^\sigma_{10;00}) Q^\sigma_{11;01} +Q^\sigma_{01;01} I^\sigma_{10;01} +\frac{4}{3} Q^\sigma_{01;01} Q^\sigma_{10;02} \right) \right. \nonumber \\
 & & \left. +\frac{2}{3} \delta_{jm} u_i q_l\ \Pi_{\sigma,\sigma} (\bm{q}) u_0^2 n_{\mathrm{imp}} I^\sigma_{10;10} \left( (I^\sigma_{01;00}+ I^\sigma_{10;00}) Q^\sigma_{11;01} \right) +I^\sigma_{01;01} Q^\sigma_{10;01} \right] \nonumber \\
 & & +\frac{e \hbar^2}{2m} u^2_0 n_{\mathrm{imp}} \int \frac{d\Omega}{2\pi} \ \Omega e^{i\Omega t} \sum_{\bm{u},\bm{q}} \sum_{i,j,k,l} e^{i(\bm{u}+\bm{q})\cdot \bm{r}} \lambda_{\mathrm{SO}}(\bm{u}) A_l(\bm{q},\Omega) \epsilon_{ijk} \sum_\sigma \delta_{\alpha z}\delta_{kz} \Re \nonumber \\
 & & \times \left[ \frac{2}{3} \delta_{jl} u_i q_m\ \Pi_{\sigma,\sigma} (\bm{u}+\bm{q}) u_0^2 n_{\mathrm{imp}} I^\sigma_{10;10} \left( 2I^\sigma_{10;00} Q^\sigma_{20;10} +Q^\sigma_{10;10} I^\sigma_{10;10} +\frac{4}{3} Q^\sigma_{10;10} Q^\sigma_{10;20} \right) \right. \nonumber \\
 & & \left. +\frac{2}{3} \delta_{jm} u_i q_l\ \Pi_{\sigma,\sigma} (\bm{q}) u_0^2 n_{\mathrm{imp}} I^\sigma_{10;10} \left( 2I^\sigma_{10;00} Q^\sigma_{20;10} +I^\sigma_{10;10} Q^\sigma_{10;10} \right) \right], \\
(j_s^\alpha)^{\mathrm{sk}}_m &\simeq& \frac{e \hbar^2}{2m} u^2_0 n_{\mathrm{imp}} \int \frac{d\Omega}{2\pi} \ \Omega e^{i\Omega t} \sum_{i,j,k} \sum_{\bm{u},\bm{q}} e^{i(\bm{u}+\bm{q})\cdot \bm{r}} \lambda_{\mathrm{SO}}(\bm{u})A_j(\bm{q},\Omega) \epsilon_{ijk} \nonumber \\
 & & \left[ \sum_\sigma (\delta_{\alpha k} \Re +\sigma \epsilon_{\alpha k} \Im) (u_i q_m+q_i u_m)\ \Pi_{\sigma, \bar{\sigma}} (\bm{u}+\bm{q}) u_0^2 n_{\mathrm{imp}} I^\sigma_{10;01} \left( I^\sigma_{00;01} I^\sigma_{10;01} + \frac{2}{3} I^\sigma_{00;01} Q^\sigma_{10;02} \right) \right. \nonumber \\
 & & \left. +\sum_\sigma \delta_{\alpha z}\delta_{kz} \Re (u_i q_m+q_i u_m)\ \Pi_{\sigma,\sigma} (\bm{u}+\bm{q}) u_0^2 n_{\mathrm{imp}} I^\sigma_{10;10} \left( I^\sigma_{00;10} I^\sigma_{10;10} + \frac{2}{3} I^\sigma_{00;10} Q^\sigma_{10;20} \right) \right].
\end{eqnarray}
Here
\begin{eqnarray}
\Pi_{\sigma, \bar{\sigma}} (\bm{u}) &\simeq & \left[ 1-\left( u_0^2 n_{\mathrm{imp}} I^\sigma_{10;01} +\hbar \Omega\ u_0^2 n_{\mathrm{imp}} I^\sigma_{10;02} +4 \frac{\hbar^2}{2m} \frac{1}{3} u^2 u_0^2 n_{\mathrm{imp}} Q^\sigma_{10;03} \right) \right]^{-1}, \\
\Pi_{\sigma,\sigma} (\bm{u}) &\simeq & \left[ 1-\left( u_0^2 n_{\mathrm{imp}} I^\sigma_{10;10} +\hbar \Omega\ u_0^2 n_{\mathrm{imp}} I^\sigma_{10;20} +4 \frac{\hbar^2}{2m} \frac{1}{3} u^2 u_0^2 n_{\mathrm{imp}} Q^\sigma_{10;30} \right) \right]^{-1}.
\end{eqnarray}
By carrying out the above integrals of the Green's functions and transforming to the real-space representation using $\bm{u} \lambda_{\mathrm{SO}}(\bm{u})=-i\int d\bm{r}'\ \nabla_{\bm{r}'} \lambda_{\mathrm{SO}}(\bm{r}') e^{-i\bm{r}'\cdot \bm{u}}$ and $\Omega q_m \bm{A}(\bm{q},\Omega)=-\int dt' \int d\bm{r}''\ \partial''_m \dot{\bm{A}}(\bm{r}'',t') e^{-i\bm{r}''\cdot \bm{q}} e^{-i\Omega t'}$, we obtain the diffusive spin current of the form 
\begin{eqnarray}
(j_s^\alpha)_m &\simeq& \frac{\pi e \hbar^2}{2m} (\nu_\uparrow+\nu_\downarrow) \nonumber \\
 & & \times \left[ (\delta_{\alpha x}+\delta_{\alpha y}) \sum_\sigma ( \bm{e}_\alpha \Re \alpha_\sigma +(\bm{e}_\alpha\times \bm{e}_z) \Im \alpha_\sigma)\cdot \int d\bm{r}' \int d\bm{r}'' \int dt' \right. \nonumber \\
 & & \ \ \ \ \ \ \ \ \int \frac{d\Omega}{2\pi} \sum_{\bm{u},\bm{q}} \frac{e^{i\Omega (t-t')+i\bm{u}\cdot (\bm{r}-\bm{r}')+i\bm{q}\cdot (\bm{r}-\bm{r}'')}}{F_\sigma -i G_\sigma} ( \nabla \lambda_{\mathrm{SO}} (\bm{r}') \times \partial_m \bm{E}_{\mathrm{em}}(\bm{r}'',t') ) \nonumber \\
 & & +\delta_{\alpha z} \sum_\sigma \alpha_\sigma \bm{e}_z \cdot \int d\bm{r}' \int d\bm{r}'' \int dt' \nonumber \\
 & & \ \ \ \ \ \ \ \ \int \frac{d\Omega}{2\pi} \sum_{\bm{u},\bm{q}} \frac{e^{i\Omega (t-t')+i\bm{u}\cdot (\bm{r}-\bm{r}')+i\bm{q}\cdot (\bm{r}-\bm{r}'')}}{(D_\sigma (\bm{u}+\bm{q})^2 -i\Omega) \tau_\sigma} ( \nabla \lambda_{\mathrm{SO}} (\bm{r}') \times \partial_m \bm{E}_{\mathrm{em}}(\bm{r}'',t') ) \nonumber \\
 & & \left. +\sum_\sigma ( \beta_{1,\sigma} (\bm{e}_i\times \bm{e}_\alpha) +\beta_{2,\sigma} ( \delta_{iz} \bm{e}_\alpha -\delta_{i\alpha} \bm{e}_z)) \cdot \nabla \lambda_{\mathrm{SO}}(\bm{r}) \int d\bm{r}' \int dt' \int \frac{d\Omega}{2\pi} \sum_{\bm{q}} \frac{e^{i\Omega (t-t')+i\bm{q}\cdot (\bm{r}-\bm{r}')}}{(D_\sigma \bm{q}^2 -i\Omega) \tau_\sigma} \nabla \cdot \bm{E}_{\mathrm{em}} (\bm{r}',t') \right], \nonumber \\
 & &
\end{eqnarray}
where
\begin{eqnarray}
\alpha_\sigma &\equiv & \left\{ \begin{array}{l}
\frac{2}{9} \frac{M \eta_+ ((1+\sigma)\eta^2_+ -M^2)}{(M^2+\eta^2_+)^3} \epsilon^2_{F\sigma} +i \frac{2}{9} \frac{\eta^2_+ ((1+\sigma)M^2 -\eta_+^2)}{(M^2+\eta^2_+)^3} \epsilon^2_{F\sigma}, \quad (\alpha=x,y) \\
-\frac{1}{9} \frac{\epsilon_{F\sigma}}{\eta_+}, \quad (\alpha=z) \\
\end{array} \right. \\
\beta_{1,\sigma} &\equiv & \left\{ \begin{array}{l}
-\frac{2}{3} \frac{\epsilon_{F\bar{\sigma}} \tau_{\bar{\sigma}}}{\hbar} \frac{M^2(\eta^2_+ -\eta_-^2)}{(M^2+\eta^2_+)(M^2+\eta^2_-)} +\pi^2 \frac{\tau_{\sigma}\nu_\sigma}{\hbar} \nu_{\uparrow} \nu_{\downarrow}\frac{\epsilon_{F\sigma}}{\eta_{\sigma}}, \quad (\alpha=x,y) \\
\pi^2 \frac{\tau_{\sigma}\nu_\sigma}{\hbar} \nu_{\uparrow} \nu_{\downarrow}\frac{\epsilon_{F\sigma}}{\eta_{\sigma}}, \quad (\alpha=z) \\
\end{array} \right. \\
\beta_{2,\sigma} &\equiv & \frac{2}{3} \frac{\epsilon_{F\bar{\sigma}} \tau_{\bar{\sigma}}}{\hbar} \frac{M (\eta_+ +\bar{\sigma}\eta_-) (\eta_+ \eta_- +\bar{\sigma} M^2)}{(M^2+\eta^2_+)(M^2+\eta^2_-)}, \\
F_\sigma &\equiv& \frac{M^2}{M^2+\eta^2_+} +\frac{2\sigma M \eta_\sigma \eta_+^2}{(M^2+\eta^2_+)^2} \Omega \tau_\sigma + \frac{\eta_\sigma^2 \eta_+^2(\eta_+^2-3M^2)}{(M^2+\eta^2_+)^3} \frac{2\epsilon_{F\bar{\sigma}} \tau_\sigma}{3m} (u+q)^2 \tau_\sigma, \\
G_\sigma &\equiv & \frac{\sigma M \eta_+}{M^2+\eta^2_+} +\frac{\eta_\sigma \eta_+ (\eta_+^2-M^2)}{(M^2+\eta^2_+)^2} \Omega \tau_\sigma +\frac{\sigma M \eta_\sigma^2 \eta_+ (3\eta_+^2-M^2)}{(M^2+\eta^2_+)^3} \frac{2\epsilon_{F\bar{\sigma}} \tau_\sigma}{3m} (u+q)^2 \tau_\sigma, \\
D_\sigma &\equiv & \frac{2 \epsilon_F \tau_\sigma}{3m}.
\end{eqnarray}

\section{Calculation of spin current driven by spin accumulation}

The spin current shown in Fig. \ref{FD} (a) and (b) reads
\begin{eqnarray}
(j_s^\alpha)_m &=& -\frac{\hbar^2}{2m} \hbar \int \frac{d\Omega}{2\pi} \ \Omega e^{i\Omega t} \sum_{i,j,k}\sum_{\bm{k},\bm{k}',\bm{u},\bm{q}} e^{i(\bm{u}+\bm{q})\cdot \bm{r}} \lambda_{\mathrm{SO}}(\bm{u}) \langle U(\bm{p})U(-\bm{p}) \rangle_{\mathrm{imp}} \mu_s(\bm{q}, \Omega) \epsilon_{ijk} \mathrm{tr}\left[ \sigma^\alpha \sigma^k \sigma^z \right] 2\Re (\chi_{\alpha ijkm}) \nonumber \\
 &=& -4i \frac{\hbar^2}{2m} \hbar u_0^2 n_{\mathrm{imp}} \int \frac{d\omega}{2\pi} \sum_{i,j}\sum_{\bm{k},\bm{k}',\bm{u},\bm{q}} e^{i(\bm{u}+\bm{q})\cdot \bm{r}} \lambda_{\mathrm{SO}}(\bm{u}) \mu_s(\bm{q}, \Omega) (\delta_{iz} \delta_{j\alpha}-\delta_{i\alpha} \delta_{jz}) \Re (\chi_{\alpha ijkm}), \\
\chi_{\alpha ijkm} &=& (k-u-k')_i (k+k')_j (2k-u-q)_m g^R_{\bm{k}} g^R_{\bm{k}'} g^R_{\bm{k}-\bm{u}} g^A_{\bm{k}-\bm{u}-\bm{q}} +(k'-k)_i (k-u+k')_j (2k-u-q)_m g^R_{\bm{k}} g^R_{\bm{k}'} g^R_{\bm{k}-\bm{u}} g^A_{\bm{k}-\bm{u}-\bm{q}} \nonumber \\
 & & +(k-u-k')_i (k+k')_j (2k-u-q)_m g^R_{\bm{k}} g^R_{\bm{k}'} g^A_{\bm{k}'-\bm{q}} g^A_{\bm{k}-\bm{u}-\bm{q}}.
\end{eqnarray}
As for the above coefficient $\chi_{\alpha ijkm}$, the contribution involving two $\bm{q}$'s and one $\bm{u}$ is calculated as follows;
\begin{eqnarray}
\chi^{(1,2)}_{\alpha ijkm} &=& -\frac{2}{3} \frac{\hbar^2}{2m} \left[ 2 k^2 u_i q_m q_j\ g^R_{\bm{k}'} (g^R_{\bm{k}})^2 (g^A_{\bm{k}})^2 +k^2 u_i q_m q_j g^R_{\bm{k}} g^R_{\bm{k}'} g^A_{\bm{k}'} (g^A_{\bm{k}})^2 +k'^2 u_i q_j q_m g^R_{\bm{k}} g^R_{\bm{k}'} (g^A_{\bm{k}'})^2 g^A_{\bm{k}} \right] \nonumber \\
 & & -\frac{8}{9} \left( \frac{\hbar^2}{2m} \right)^2 k^2 k'^2 q_j u_m q_i\ g^R_{\bm{k}'} (g^A_{\bm{k}'})^2 g^R_{\bm{k}} (g^A_{\bm{k}})^2.
\end{eqnarray}
The resultant spin current reduces to
\begin{eqnarray}
(j_s^\alpha )^{(1,2)}_m &=& -4 \frac{\hbar^2}{2m} \hbar u_0^2 n_{\mathrm{imp}} \int \frac{d\Omega}{2\pi} \ \Omega e^{i\Omega t} \sum_{\bm{u},\bm{q}} e^{i(\bm{u}+\bm{q})\cdot \bm{r}} \lambda_{\mathrm{SO}}(\bm{u}) \mu_s(\bm{q}, \Omega) (u_\alpha q_m q_z -u_z q_m q_\alpha)\ \frac{4}{3} \Re \left[ I^\sigma_{10;00} Q^\sigma_{20;20} + I^\sigma_{10;10} Q^\sigma_{10;20} \right]. \nonumber \\
\end{eqnarray}
Now we perform the vertex correction $\chi_{\alpha ijkm} \mathrm{tr}\left[ \sigma^\alpha \sigma^k \sigma^z \right] \to \tilde{\chi}_{\alpha ijk} \sum_{a,b,c,d,e} \sigma^\alpha_{da} \Gamma_{ad,cb} \sigma^k_{be} \sigma^z_{ec}$ with $\Gamma_{ad,cb}$ being the vertex correction corresponding to the ladder diagram including the normal impurity and the SOC due to impurities \cite{Hosono}, and
\begin{eqnarray}
\tilde{\chi}_{\alpha ijk} &=& (k-u-k')_i (k+k')_j g^R_{\bm{k}} g^R_{\bm{k}'} g^R_{\bm{k}-\bm{u}} g^A_{\bm{k}-\bm{u}-\bm{q}} +(k'-k)_i (k-u+k')_j g^R_{\bm{k}} g^R_{\bm{k}'} g^R_{\bm{k}-\bm{u}} g^A_{\bm{k}-\bm{u}-\bm{q}} \nonumber \\
 & & +(k-u-k')_i (k+k')_j g^R_{\bm{k}} g^R_{\bm{k}'} g^A_{\bm{k}'-\bm{q}} g^A_{\bm{k}-\bm{u}-\bm{q}}.
\end{eqnarray}
According to Ref. \onlinecite{Hosono}, the vertex correction is calculated as 
$
\Gamma_{ad,cb}(\bm{q},\Omega)=\Gamma_C(\bm{q},\Omega) \delta_{ad} \delta_{cb}+\Gamma_S(\bm{q},\Omega) \sum_l \sigma^l_{ad} \sigma^l_{cb}
$
with
\begin{eqnarray}
\Gamma_C(\bm{q},\Omega) &=& \frac{1+3\kappa}{2(Dq^2+i\Omega)\tau}, \\
\Gamma_S(\bm{q},\Omega) &=& \frac{(1-\kappa)(1+3\kappa)}{2(4\kappa+(Dq^2+i\Omega)\tau)},
\end{eqnarray}
leading to
\begin{eqnarray}
\sum_{a,b,c,d,e} \sigma^\alpha_{da} \Gamma_{ad,cb} \sigma^k_{be} \sigma^z_{ec} &=& \Gamma_C \mathrm{tr}[\sigma^\alpha \hat{1}] \mathrm{tr}[\hat{1} \sigma^k \sigma^z] +\sum_l \Gamma_S \mathrm{tr}[\sigma^\alpha \sigma^l] \mathrm{tr}[\sigma^l \sigma^k \sigma^z] \nonumber \\
 &=& 4i \epsilon_{\alpha k z} \Gamma_S.
\end{eqnarray}
Here, $\tau\equiv \hbar/2\pi u_0^2 n_{\mathrm{imp}} \nu$, $\kappa\equiv \lambda_{\mathrm{SO}}^2 k_F^4/3$ ($k_F$ is the Fermi wave number), $D\equiv 2\epsilon_F \tau/3m$, and $\xi\equiv \sqrt{4\kappa/D\tau}$.
$\hat{1}$ is the identity matrix.
Therefore, we obtain the spin current including the vertex correction of the form
\begin{eqnarray}
(\tilde{j_s^\alpha})^{(1,2)}_m &\simeq & \frac{4\pi \hbar^2 \tau \nu}{3m} \int d\bm{r}' \int d\bm{r}'' \int dt'\ \int \frac{d\Omega}{2\pi} \sum_{\bm{u},\bm{q}} e^{i\Omega (t-t')+i\bm{u}\cdot (\bm{r}-\bm{r}')+i\bm{q}\cdot (\bm{r}-\bm{r}'')} \frac{\Omega/\tau (1-\kappa) (1+3\kappa)}{D((\bm{u}+\bm{q})^2+\xi^2) +i\Omega} \nonumber \\
 & & \times \left( \partial_\alpha \lambda_{\mathrm{SO}}(\bm{r}')\ \partial_z (\partial_m \mu_s(\bm{r}'', t')) -\partial_z \lambda_{\mathrm{SO}}(\bm{r}')\ \partial_\alpha (\partial_m \mu_s(\bm{r}'', t')) \right),
\end{eqnarray}
with $\nu$ being the density of state.
We remark that the spin-diffusion propagator is given by 
\begin{eqnarray}
D(\bm{r}_1,\bm{r}_2,t) &:=& \int \frac{d\Omega}{2\pi} \sum_{\bm{u},\bm{q}} \frac{e^{i\Omega t+i\bm{u}\cdot \bm{r}_1+i\bm{q}\cdot \bm{r}_2} \Omega/\tau (1-\kappa) (1+3\kappa)}{D((\bm{u}+\bm{q})^2+\xi^2) +i\Omega} \nonumber \\
 &=& -\delta(\bm{r}_2-\bm{r}_1) \sqrt{\frac{\pi}{Dt}} \left( D\xi^2-\frac{|\bm{r}_1|^2}{4Dt^2}+\frac{1}{2t}\right) \exp\left[ -\left( D\xi^2 t+\frac{|\bm{r}_1|^2}{4Dt} \right)\right].
\end{eqnarray}
On the other hand, the contribution involving two $\bm{u}$'s and one $\bm{q}$ is given as follows;
\begin{eqnarray}
\chi^{(2,1)}_{\alpha ijkm} &=& -\frac{4}{3} \frac{\hbar^2}{2m} k^2 u_i u_j q_m [ g^R_{\bm{k}} g^R_{\bm{k}'} g^R_{\bm{k}})^2 g^A_{\bm{k}} -\frac{8}{15} \left( \frac{\hbar^2}{2m} \right)^2 k^4 (u_i u_j q_m +u_i u_o q_o \delta_{jm} +u_i u_m q_j) g^R_{\bm{k}} g^R_{\bm{k}'} (g^R_{\bm{k}})^2 (g^A_{\bm{k}})^2 \nonumber \\
 & & -\frac{4}{3} \frac{\hbar^2}{2m} k^2 u_i (u_m q_j+q_m u_j) g^R_{\bm{k}} g^R_{\bm{k}'} g^R_{\bm{k}} (g^A_{\bm{k}})^2 -\frac{1}{3}k^2 (u_iu_mq_j+u_iu_jq_m) g^R_{\bm{k}} g^R_{\bm{k}'} g^A_{\bm{k}'} (g^A_{\bm{k}})^2 \nonumber \\
 & & -\frac{1}{3}k'^2 u_iu_mq_j g^R_{\bm{k}} g^R_{\bm{k}'} (g^A_{\bm{k}'})^2 g^A_{\bm{k}} -\frac{4}{9}k^2k'^2 u_iu_mq_j g^R_{\bm{k}} g^R_{\bm{k}'} (g^A_{\bm{k}'})^2 (g^A_{\bm{k}})^2.
\end{eqnarray}
The corresponding spin current reduces to
\begin{eqnarray}
(j_s^\alpha)^{(2,1)}_m &=& -4 \frac{\hbar^2}{2m} \hbar u_0^2 n_{\mathrm{imp}} \int \frac{d\Omega}{2\pi}\ \Omega e^{i\Omega t} \sum_{\bm{u},\bm{q}} e^{i(\bm{u}+\bm{q})\cdot \bm{r}} \lambda_{\mathrm{SO}}(\bm{u}) \mu_s(\bm{q},\Omega) \sum_{i,j,k} \epsilon_{ijk} \epsilon_{\alpha k z} \nonumber \\
 & & \times \left( u_iu_mq_j \Re \left[ -\frac{4}{3}I^\sigma_{10;00} Q^\sigma_{20;20}-\frac{4}{3}I^\sigma_{10;10} Q^\sigma_{10;20}-\frac{16}{9}(Q^\sigma_{10;20})^2-\frac{16}{15}I^\sigma_{10;00} S^\sigma_{30;20} \right] +u_iu_oq_o\delta_{jm} \Re \left[-\frac{16}{15}I^\sigma_{10;00} S^\sigma_{30;20} \right] \right). \nonumber \\
\end{eqnarray}
By performing the vertex correction, we obtain the final expression of the spin current
\begin{eqnarray}
(\tilde{j_s^\alpha})^{(2,1)}_m &\simeq & \frac{152\pi \hbar^2 \tau \nu}{45m}\left( \frac{2\epsilon_F \tau}{\hbar} \right)^2 \int d\bm{r}' \int d\bm{r}'' \int dt'\ \int \frac{d\Omega}{2\pi} \sum_{\bm{u},\bm{q}} e^{i\Omega (t-t')+i\bm{u}\cdot (\bm{r}-\bm{r}')+i\bm{q}\cdot (\bm{r}-\bm{r}'')} \frac{\Omega/\tau (1-\kappa) (1+3\kappa)}{D((\bm{u}+\bm{q})^2+\xi^2) +i\Omega} \nonumber \\
 & & \times \left( \partial_\alpha \mu_s(\bm{r}', t')\ \partial_z (\partial_m \lambda_{\mathrm{SO}}(\bm{r}'')) -\partial_z \mu_s(\bm{r}', t')\ \partial_\alpha (\partial_m \lambda_{\mathrm{SO}}(\bm{r}'')) \right).
\end{eqnarray}

\end{document}